\documentclass[11pt]{article}
\usepackage{amssymb}
\usepackage{epsfig}
\usepackage{palatino}

\newcommand{\qedsymb}{\hfill{\rule{2mm}{2mm}}}

\def\mod{{\rm{mod}}}
\def\poly{{\rm{poly}}}

\newcommand\ip[1]{{\langle {#1} \rangle}}

\newcommand\ket[1]{{ |{#1} \rangle }}

\renewcommand{\epsilon}{\varepsilon}

\begin{document}

\title{\bf A Subexponential Time Algorithm for the Dihedral Hidden Subgroup Problem with Polynomial Space}

\author{
 Oded Regev \footnote{Department of Computer Science, Tel-Aviv University, Tel-Aviv 69978, Israel. Work
   supported by an Alon Fellowship and the Army Research Office grant DAAD19-03-1-0082.} }


\maketitle

\begin{abstract}
In a recent paper, Kuperberg described the first subexponential time algorithm for solving the dihedral hidden subgroup
problem. The space requirement of his algorithm is super-polynomial. We describe a modified algorithm whose running
time is still subexponential and whose space requirement is only polynomial.
\end{abstract}

\section{Introduction}

A central problem in quantum computation is the hidden subgroup problem (HSP). Here, we are given a black box that
computes a function on elements of a group $G$. The function is known to be constant and distinct on left cosets of a
subgroup $H\subseteq G$ and our goal is to find $H$. Interestingly, most known quantum algorithms that provide a
super-polynomial advantage over classical algorithms solve special cases of the HSP on Abelian groups. There has also
been considerable interest in the HSP on noncommutative groups (see, e.g.,
\cite{GrigniSchulman01,HallgrenTashma00,RottelerWreathGroup,SanthaHiddenTranslation}). For example, one important group
is the symmetric group: it is known that solving the HSP on the symmetric group leads to a solution to graph
isomorphism~\cite{KoblerGraphIsomorphism}.

In this paper we will be interested in the HSP on the dihedral group. The dihedral group of order $2N$, denoted $D_N$,
is the group of symmetries of an $N$-sided regular polygon. It consists of $N$ rotations, which we denote by
$(0,0),\ldots,(0,N-1)$, and $N$ reflections, which we denote by $(1,0),\ldots,(1,N-1)$. It is isomorphic to the
abstract group generated by the element $\rho$ of order $n$ and the element $\tau$ of order 2 subject to the relation
$\rho\tau = \tau \rho^{-1}$. Regev~\cite{Regev02B} showed that under certain conditions, an efficient solution to the
dihedral HSP implies a quantum algorithm for lattice problems. This gives a strong incentive to finding an efficient
solution to the dihedral HSP.

However, although the dihedral group is one of the simplest noncommutative groups, no efficient solution to the
dihedral HSP is known. Ettinger and H{\o}yer~\cite{EttingerHoyerDihedral} showed that one can obtain sufficient
statistical {\em information} about the hidden subgroup with only a polynomial number of queries to the black box.
However, there is no efficient algorithm that solves the HSP using this information. In fact, it was shown in
\cite{Regev03A} that solving HSP using this information is a hard problem in a certain precise sense.

Recently, Kuperberg \cite{Kuperberg} presented the first subexponential time algorithm for the dihedral HSP. Namely,
his algorithm runs in time $2^{O(\sqrt{\log{N}})}$ (the input size is $O(\log N)$). This is currently the best known
algorithm for the dihedral HSP. However, in order to achieve this running time, Kuperberg's algorithm requires
$2^{O(\sqrt{\log{N}})}$ space. Essentially, this happens since the algorithm keeps many qubits around until certain
collisions occur. Our main result in this paper is an algorithm that requires only polynomial space, i.e., $\poly(\log
N)$. The running time of our algorithm is still subexponential and only slightly higher than Kuperberg's algorithm,
namely, $2^{O(\sqrt{\log{N}\log\log{N}})}$.

Our algorithm combines ideas from Kuperberg's algorithm \cite{Kuperberg} and a paper by Regev \cite{Regev02B}. Our
classical abstraction of the problem is influenced by a paper by Blum, Kalai and Wasserman \cite{BlumKW}. We start in
Section \ref{sec:kup} with a simplified description of Kuperberg's algorithm. Then, in Section \ref{sec:new}, we
describe our new algorithm.

\section{Kuperberg's Algorithm}\label{sec:kup}

In this section we present a simplified description of Kuperberg's algorithm. We concentrate on the basic idea and try
to omit some of the more technical issues. We start with describing an algorithm for a certain classical problem. We
will later show that this algorithm corresponds exactly to Kuperberg's algorithm.

\subsection{A Classical Abstraction}

For simplicity, we only consider the case where $N=2^n$ and $n=k^2+1$ for some integer $k$. The algorithm can be
modified to work without this assumption.

Let us consider the following classical scenario. We are dealing with `objects' that are labelled with numbers modulo
$N=2^n$ (eventually, these objects will turn out to be qubits, but let's forget about that for now). Our goal is to
obtain an object whose label is $2^{n-1}$. These objects are created by a `machine' that we have at our disposal. This
machine outputs both an object and its label. We are guaranteed that the machine outputs objects whose label is chosen
uniformly at random from $\{0,\ldots,2^n-1\}$. Each time we ask the machine for a new object we pay one time unit. So
here is our first algorithm: call the machine repeatedly until it happens to output an object whose label is $2^{n-1}$.
Clearly, this algorithm requires $O(2^n)$ time units on average.

It turns out that these objects have a nice property: given two objects, labelled with $a$ and $b$, we can {\em
combine} them and obtain a new object whose label is $a-b$ (the two original objects are gone). This combination
operation succeeds with probability 50\%; with probability 50\%, the operation fails and then both original objects are
gone. Let us now show how to obtain an algorithm whose running time is $2^{O(\sqrt{n})}$. This is the basic idea
underlying \cite{Kuperberg}.

\begin{figure}[h]
\center{
 \epsfxsize=4in
 \epsfbox{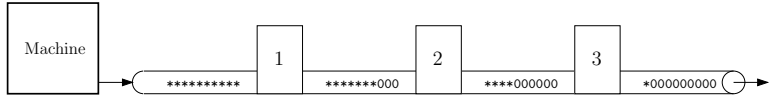}}
 \caption{A pipeline of routines with $n=10$, $k=3$.}
 \label{fig:pipeline}
\end{figure}

The overall structure of the algorithm is that of a `pipeline' of $k$ routines, as in Figure \ref{fig:pipeline}. That
is, the input to routine $i+1$ is the output of the routine $i$. The input to routine $1$ are `fresh' objects from the
machine, i.e., objects whose labels are chosen uniformly at random. For any $i=1,\ldots,k$, the output of routine $i$
(and the input to routine $i+1$) are objects whose labels have the following distribution: the $ik$ least significant
bits equal 0 and the remaining $n-ik$ bits are chosen uniformly at random. In other words, each routine is supposed to
output objects whose labels have $k$ additional bits zeroed out. Notice that with probability 50\%, the last routine
(i.e., routine $k$) outputs an object whose label is $2^{n-1}$.

It remains to describe how to implement the routines. Let us describe routine $i$ for some $i=1,\ldots,k$. The routine
maintains a pile of objects. Initially, the pile is empty. Whenever a new object arrives, the routine compares the $k$
bits in positions $(i-1)k+1,\ldots,ik$ of its label to the same bits in all of the objects currently in its pile
(notice that the bits in positions $1,\ldots,(i-1)k$ are guaranteed to be zero). If no match is found (i.e., no object
currently in the pile has the same setting to these $k$ bits) then the routine adds the new object to the pile. If a
match is found then the routine {\em combines} the new object with the matching object in the pile. With probability
50\%, the combination is successful and the machine outputs the resulting object. Notice that the $ik$ least
significant bits of the label of the resulting object are all zero. Moreover, the remaining $n-ik$ bits are still
random since the behavior of the routine does not depend on them.

Finally, let us show that the expected running time of the algorithm is indeed $2^{O(\sqrt{n})}$. In other words, we
will show that this is the amount of time it takes to obtain one object from the last routine in the pipeline. The
intuitive idea is the following. Initially, the pile of a routine is empty and matches rarely occur. However, after
around $2^k$ objects the pile gets rather full; from that point on, the routine needs an average of four objects in
order to produce one output object (since we are combining two objects in order to produce one output object and our
success probability is 50\%). Hence, the number of objects needed from the machine in order to produce one object by
the final routine is roughly $4^k$.

Let us make this argument more formal. First we observe that with very high probability, a routine that gets as input
$l\cdot 2^k$ objects for some $l\ge 8$, outputs at least $l/8 \cdot 2^k$ objects. This follows by noting that at most
$2^k$ of these objects can remain in the pile. On the remaining $(l-1)\cdot 2^k$ objects, the routine performs
combination operations. The expected number of output objects is therefore $(l-1)/4 \cdot 2^k$; a simple application of
the Chernoff bound shows that with very high probability, the number of output objects is at least $l/8 \cdot 2^k$. We
can now complete the proof by noting that if the first routine is given $8^k\cdot 2^k = 2^{O(\sqrt{n})}$ objects then
with very high probability the last routine outputs at least one object (and in fact, at least $2^k$ objects).

\subsection{The Quantum Setting}

We now show how to obtain from the above an algorithm for the dihedral hidden subgroup problem. We are given oracle
access to a function $f:D_N \rightarrow R$ from the dihedral group to some arbitrary set $R$. The function is promised
to be constant on cosets of some subgroup $H \subseteq D_N$ and distinct on different cosets. Our goal is to extract
the subgroup $H$. Ettinger and H{\o}yer \cite{EttingerHoyerDihedral} showed that it is enough to solve the problem for
the case where $H=\{(0,0),(1,d)\}$ is generated by a reflection $(1,d)$. Hence, our goal now is to find $d$, a number
between $0$ and $2^n-1$.

In fact, finding the {\em least significant bit} of $d$ is enough. Indeed, let us show how to find $d$ given an
algorithm that only finds the least significant bit of $d$. We start by calling the algorithm once with the given
oracle. This allows us to obtain the least significant bit of $d$. Assume the answer is `0'. Then, consider the
function $f':D_{N/2} \rightarrow R$ given by $f'(a,b):=f(a,2b)$. Notice that this function hides the subgroup
$\{(0,0),(1,d/2)\}$ of $D_{N/2}$. Similarly, if the answer is `1', consider the function $f'':D_{N/2} \rightarrow R$
given by $f''(a,b):=f(a,2b+1)$. This function hides the subgroup $\{(0,0),(1,(d-1)/2)\}$ of $D_{N/2}$. We can now
obtain the second least significant bit of $d$ by calling the algorithm with either $f'$ or $f''$. By continuing this
process, we can find all the bits of $d$.

Hence, in the following we show how to obtain the least significant bit of $d$. We start with a simple quantum routine
that produces certain one-qubit states. First, we create the uniform superposition over all elements of $D_N$. Namely,
we create the state
$$ \sum_{b,x}\ket{b,x}$$
where $b$ ranges over $\{0,1\}$ and $x$ ranges over $\{0,\ldots,N-1\}$. Here and in the following we omit the
normalizing factor.
 We now add some qubits and call the oracle. The resulting state is
$$ \sum_{b,x}\ket{b,x}\ket{f(b,x)}.$$
After measuring the last register, the state collapses to
$$ \ket{0,x} + \ket{1,x+d ~\mod~ N} $$
for some arbitrary $x$. We perform a standard (Abelian) Fourier transform on the second register and obtain
$$ \sum_{y=0}^{N-1} \exp(2\pi i xy / N) \ket{0,y} + \sum_{y=0}^{N-1} \exp(2\pi i (x+d)y / N) \ket{1,y}.$$
Finally, we measure $y$ and obtain the one-qubit state
$$ \ket{0} + \exp(2\pi i d y / N) \ket{1}.$$
Notice that $y$ is distributed uniformly on $\{0,1,\ldots,N-1\}$ and is known to us.

The above routine can be seen as the `machine' in the classical scenario described above. Namely, an object with label
$y \in \{0,\ldots,2^n-1\}$ is simply the one-qubit state $ \ket{0} + \exp(2\pi i d y / N) \ket{1}$. Combining two
objects is done as follows. Given $ \ket{0} + \exp(2\pi i d y_1 / N) \ket{1}$ and $ \ket{0} + \exp(2\pi i d y_2 / N)
\ket{1}$ we tensor them together and obtain
 $$\ket{00} + \exp(2\pi i d y_1 / N) \ket{10} + \exp(2\pi i d y_2 / N) \ket{01} + \exp(2\pi i d (y_1+y_2) / N)
 \ket{11}$$
We now measure the parity of the two qubits. With probability 50\%, we measure `odd' and the state collapses to
 $$\exp(2\pi i d y_1 / N) \ket{10} + \exp(2\pi i d y_2 / N) \ket{01}.$$
By omitting the global phase and renaming the basis states, this is equivalent to
 $$\ket{0} + \exp(2\pi i d (y_2-y_1) / N) \ket{1},$$
as required. Hence, we can apply the algorithm described above and obtain, after $2^{O(\sqrt{n})}$ operations, the
state
 $$\ket{0} + \exp(2\pi i d 2^{n-1} / N) \ket{1} = \ket{0} + \exp(\pi i d) \ket{1}.$$
Measuring this state in the Hadamard basis yields the least significant bit of $d$.

\section{A Polynomial Space Algorithm}\label{sec:new}

In this section we present our new algorithm. As can be seen from the above description, each routine has to store
$\approx 2^{\sqrt{n}}$ objects (i.e., qubits) before a collision is found. Hence, the space requirement is
$2^{O(\sqrt{n})}$. The space requirement of the algorithm we present in this section is only polynomial. The running
time is only slightly larger, namely, $2^{O(\sqrt{n \log n})}$.

For simplicity, assume that $n=1+kl$ where $k=O(\sqrt{n/\log n})$ and $l=O(\sqrt{n \log n})$ are both integer (we could
also take $n=1+k^2$ as before but this would lead to a slightly worse running time). Our algorithm is based on a
different combination operation. This operation takes as input $l+4$ labelled objects whose labels are uniformly
distributed and with constant probability outputs one object that has its $l$ least significant bits zeroed out. This
operation is performed as follows. Assume our input is
$$ \ket{0} + \exp(2\pi i \cdot d y_j / N) \ket{1}, \quad j=1,\ldots,l+4.$$
We tensor together all these qubits and obtain
$$ \sum_{\vec{b} \in \{0,1\}^{l+4}} \exp(2\pi i \cdot d \cdot \ip{\vec{b}, \vec{y}}/ N) \ket{\vec{b}}$$
where $\vec{y}$ denotes $(y_1,\ldots,y_{l+4})$ and $\ip{\vec{b}, \vec{y}}$ denotes $\sum_j b_j y_j$. Since we know
$y_1,\ldots,y_l$ we can compute $\ip{\vec{b},\vec{y}} ~\mod~ 2^l$ in an extra register and obtain
$$ \sum_{\vec{b} \in \{0,1\}^{l+4}} \exp(2\pi i \cdot d \cdot \ip{\vec{b}, \vec{y}} / N) \ket{\vec{b}} \ket{\ip{\vec{b},\vec{y}} ~\mod~ 2^l}.$$
We now measure the second register and obtain some value $z \in \{0,\ldots,2^l-1\}$. We then compute (classically) the
number $m$ of bit strings $\vec{b} \in \{0,1\}^{l+4}$ for which $\ip{\vec{b},\vec{y}} ~\mod~ 2^l = z$. This is done in
a brute-force way and hence takes time $O(2^l)$. If $m$ is less than two or more than, say, 32, then we say that the
combination operation failed. Otherwise, the state that we have is
$$ \sum_{j=1}^m \exp(2\pi i \cdot d \cdot \ip{\vec{b}^j, \vec{y}} / N) \ket{\vec{b}^j}$$
where $\vec{b}^1,\ldots,\vec{b}^m \in \{0,1\}^{l+4}$ are the bit strings that we found. We would like to remain with
exactly two terms in the above sum. So we perform a projective measurement on the subspace spanned by $\ket{\vec{b}^1}$
and $\ket{\vec{b}^2}$ (we can do this since we know the $\vec{b}^j$'s). With constant probability we have the state
$$ \exp(2\pi i \cdot d \cdot \ip{\vec{b}^1, \vec{y}} / N) \ket{\vec{b}^1} + \exp(2\pi i \cdot d \cdot \ip{\vec{b}^2, \vec{y}} / N) \ket{\vec{b}^2}.$$
By omitting the global phase and renaming, we obtain the one-qubit state
$$ \ket{0} + \exp(2\pi i \cdot d \cdot \ip{\vec{b}^2-\vec{b}^1, \vec{y}} / N) \ket{1}.$$
This is exactly the object whose label is $\ip{\vec{b}^2-\vec{b}^1, \vec{y}}$. Since $\ip{\vec{b}^1, \vec{y}} ~\mod~
2^l = \ip{\vec{b}^2, \vec{y}} ~\mod~ 2^l = z$, the $l$ least significant bits of this label are all zero.

It remains to show why the event that $m \in \{2,3,\ldots,32\}$ happens with constant probability over the choice of
$\vec{y}$. Fix some $z \in \{0,\ldots,2^{l}-1\}$. For each $\vec{b} \in \{0,1\}^{l+4} \setminus \{0^{l+4}\}$ we define
an indicator random variable $X_{\vec{b}}$ that is $1$ if $\ip{\vec{b},\vec{y}} ~\mod~ 2^l = z$ and $0$ otherwise (for
convenience we ignore the all zero string since $\ip{0^{l+4},\vec{y}}$ is always zero). Each random variable has
expected value $2^{-l}$ and variance $2^{-l} - 2^{-2l}$. These random variables are pairwise independent. Let $Y=\sum
X_{\vec{b}}$ over all $\vec{b} \in \{0,1\}^{l+4} \setminus \{0^{l+4}\}$. Its expected value is $16 - 2^{-l} \approx
16$. Its variance is $16 - 17 \cdot 2^{-l} + 2^{-2l} \approx 16$. By Chebyshev's inequality we obtain that $Y \in
\{2,3,\ldots,32\}$ with some constant probability. Hence, the expected fraction of $z$'s that have this number of
$\vec{b}$'s mapped to them is constant. Hence, the above procedure is successful with constant probability.

We note that a similar combination operation can be performed on other $l$-bit blocks. For example, given $l+4$ objects
whose labels have their $l$ least significant bits all zero and the next $l$ bits (from location $l+1$ to $2l$) are
uniformly distributed, we can extract with constant probability one object such that its label has its $2l$ least
significant bits zero.

Using this combination operation, we can now describe our new algorithm. As before, the algorithm operates as a
pipeline of $k$ routines. The output of routine $i$ consists of objects whose labels have their $il$ least significant
bits zeroed out. Unlike the previous algorithm, there is no need for a pile. Each routine simply waits until it
receives $l+4$ objects from the previous routine and then it uses the combination operation to obtain one object with
$l$ additional bits zeroed out. Recall that with constant probability the combination operation is successful. By using
the Chernoff bound, one can show that if we input $l^{O(k)}=2^{O(\sqrt{n \log n})}$ objects to the pipeline then with
very high probability, the last routine outputs at least one object. Each combination operation takes
$2^{O(l)}=2^{O(\sqrt{n \log n})}$ time and hence the total running time is also $2^{O(\sqrt{n \log n})}$.

\section{Acknowledgments}

I would like to thank G{\'a}bor Ivanyos, Julia Kempe, and Miklos Santha for useful discussions.


\begin{thebibliography}{10}

\bibitem{BlumKW}
A.~Blum, A.~Kalai, and H.~Wasserman.
\newblock Noise-tolerant learning, the parity problem, and the statistical
  query model.
\newblock {\em J. ACM}, 50(4):506--519, 2003.
\newblock Preliminary version in STOC'00.

\bibitem{EttingerHoyerDihedral}
M.~Ettinger and P.~H{\o}yer.
\newblock On quantum algorithms for noncommutative hidden subgroups.
\newblock {\em Adv. in Appl. Math.}, 25(3):239--251, 2000.

\bibitem{SanthaHiddenTranslation}
K.~Friedl, G.~Ivanyos, F.~Magniez, M.~Santha, and P.~Sen.
\newblock Hidden translation and orbit coset in quantum computing.
\newblock In {\em Proc. 35th ACM Symp. on Theory of Computing}, 2003.

\bibitem{GrigniSchulman01}
M.~Grigni, L.~J. Schulman, M.~Vazirani, and U.~V. Vazirani.
\newblock Quantum mechanical algorithms for the nonabelian hidden subgroup
  problem.
\newblock In {\em Proc. 33rd ACM Symp. on Theory of Computing}, pages 68--74,
  2001.

\bibitem{HallgrenTashma00}
S.~Hallgren, A.~Russell, and A.~Ta-Shma.
\newblock Normal subgroup reconstruction and quantum computation using group
  representations.
\newblock In {\em Proc. 32nd ACM Symp. on Theory of Computing}, pages 627--635,
  2000.

\bibitem{KoblerGraphIsomorphism}
K.~Johannes, S.~Uwe, and T.~Jacobo.
\newblock {\em The graph isomorphism problem: its structural complexity}.
\newblock Birkh\"auser Boston Inc., 1993.

\bibitem{Kuperberg}
G.~Kuperberg.
\newblock A subexponential-time quantum algorithm for the dihedral hidden
  subgroup problem.
\newblock In {\em quant-ph/0302112, http://xxx.lanl.gov}, 2003.

\bibitem{Regev03A}
O.~Regev.
\newblock New lattice based cryptographic constructions.
\newblock In {\em Proc. 35th ACM Symp. on Theory of Computing (STOC)}, pages
  407--416, 2003.

\bibitem{Regev02B}
O.~Regev.
\newblock Quantum computation and lattice problems.
\newblock {\em SIAM Journal on Computing}, 33(3):738--760, 2004.
\newblock Preliminary version in FOCS'02.

\bibitem{RottelerWreathGroup}
M.~{R\"otteler} and T.~Beth.
\newblock Polynomial-time solution to the hidden subgroup problem for a class
  of non-abelian groups.
\newblock In {\em quant-ph/9812070, http://xxx.lanl.gov}, 1998.

\end{thebibliography}
\end{document}